\documentclass[conference]{IEEEtran}
\usepackage{cite}
\usepackage{amsmath,amssymb,amsfonts}
\usepackage[caption=false]{subfig}
\usepackage{bm}
\usepackage{algorithm2e}
\usepackage{enumitem}
\usepackage{tabularx}
\usepackage{makecell}
\usepackage{graphicx}
\usepackage{textcomp}
\usepackage{multirow}
\usepackage{lipsum}

\DeclareMathOperator*{\argmin}{arg\,min}
\usepackage{hyperref}
\hypersetup{colorlinks,allcolors=black}

\newcolumntype{L}[1]{>{\raggedright\let\newline\\\arraybackslash\hspace{0pt}}p{#1}}		
\newcolumntype{C}[1]{>{\centering\let\newline\\\arraybackslash\hspace{0pt}}p{#1}}
\newcolumntype{R}[1]{>{\raggedleft\let\newline\\\arraybackslash\hspace{0pt}}p{#1}}

\begin{document} 

\title{
Online Slice Reconfiguration for End-to-End QoE in 6G Applications
}

\author{
\IEEEauthorblockN{
Dibbendu~Roy\IEEEauthorrefmark{1}, 
Aravinda~S.~Rao\IEEEauthorrefmark{1}, 
Tansu~Alpcan\IEEEauthorrefmark{1}, 
Akilan Wick\IEEEauthorrefmark{2},
Goutam~Das\IEEEauthorrefmark{3},
Marimuthu~Palaniswami\IEEEauthorrefmark{1}}
\IEEEauthorblockA{\IEEEauthorrefmark{1}Department of Electrical and Electronic Engineering, The University of Melbourne, Parkville, VIC - 3010, Australia. \\
Email: dibbendu.roy@student.unimelb.edu.au, \{aravinda.rao, tansu.alpcan, palani\}@unimelb.edu.au}
\IEEEauthorblockA{\IEEEauthorrefmark{2} Mobile Development and Product Engineering, Telstra, Melbourne, VIC - 3000, Australia \\
Email: Akilan.Wick@team.telstra.com}
\IEEEauthorblockA{\IEEEauthorrefmark{3} G.S. Sanyal School of Telecommunications, Indian Institute of Technology Kharagpur, West Bengal 721302, India.\\
Email: gdas@gssst.iitkgp.ac.in}
}

\maketitle

\begin{abstract}
End-to-end (E2E) quality of experience (QoE) for 6G applications depends on the synchronous allocation of networking and computing resources, also known as slicing. However, the relationship between the resources and the E2E QoE outcomes is typically stochastic and non-stationary. Existing works consider known resource demands for slicing and formulate optimization problems for slice reconfiguration. In this work, we create and manage slices by learning the relationship between E2E QoE and resources. We develop a gradient-based online slice reconfiguration algorithm (OSRA) to reconfigure and manage slices in resource-constrained scenarios for radio access networks (RAN). We observe that our methodology meets the QoE requirements with high accuracy compared to existing approaches. It improves upon the existing approaches by approximately 98\% for bursty traffic variations. Our algorithm has fast convergence and achieves low E2E delay violations for lower priority slices. 
\end{abstract}

\begin{IEEEkeywords}
E2E QoE, Network Slicing, Kubernetes, SDN, O-RAN.
\end{IEEEkeywords}

\section{Introduction}\label{sec:introduction}
The International Telecommunication Union (ITU) had classified 5G mobile network services into three broad categories based on their end-to-end (E2E) quality of experience (QoE) requirements (metrics such as delay, throughput, jitter, and reliability): Enhanced Mobile Broadband (bandwidth-intensive like Virtual Reality), Ultra-reliable and Low-latency Communications (highly delay-sensitive and reliable like automated driving), and Massive Machine Type Communications (high connection density like IoT and Industry 4.0) \cite{sachs20185g}. In 6G, further granularity might redefine these categories \cite{letaief2019roadmap}. 6G thrives towards zero-touch E2E networks \cite{Yourguid58:online} to serve these modern applications. This requires integration of intelligence in the network for automated management of resources without user intervention. To facilitate the same, the concept of \textit{slicing} was introduced which virtualizes network and computing resources over the same physical infrastructure. This virtualization is facilitated by modern technologies such as \textit{software defined networks (SDN)}  \cite{chartsias2017sdn} at the network side, \textit{Kubernetes} (K8) \cite{kubernetes_2021} and \textit{dockers} \cite{docker} at the servers. 


Based on the QoE requirements of the applications, we can classify the applications into different classes. The virtualization allows each application class to have a \textit{slice}. 
Therefore, the service providers must jointly design and adapt their slices to satisfy the diverse end-to-end (E2E) Quality of Experience (QoE). However, the relationship between the E2E QoE parameters and the resources is non-linear (typically non-convex), stochastic and non-stationary \cite{she2019cross}. Thus, in order to manage the slices to ensure E2E QoE requirements, intelligent dynamic learning of the relationship between QoE parameters and resources becomes necessary. 

Creating network slices and managing them is an active area of research \cite{wang2018fast,wei2020dynamic,rossi2020hierarchical,hirai2020automated}. These works however assume the resource demands for the applications to be known. This assumption is quite restrictive and does not serve the purpose of meeting E2E QoE. In addition, to the best of our knowledge, none of the existing works model both SDN and Kubernetes to jointly create E2E slice. Thus, service providers should facilitate E2E slice creation and management in an online, automated and intelligent manner.
This paper presents a novel joint optimization algorithm for slice management that serves the mentioned purpose. We consider the fronthaul segment of the radio access network (RAN) \cite{Microsof16:online}, where the traffic patterns are usually bursty. 
In particular, due to the dynamic nature of application arrivals, the slices might require re-configuration to maintain E2E QoE. In this paper, we design an algorithm to facilitate this re-configuration in a fast and efficient manner in resource constrained and unconstrained scenarios.
Our contributions are:
\begin{itemize}
    \item We create and manage slices without assuming the knowledge of required resources.
    \item We capture true E2E experience by creating E2E slices with joint allocation of networking and computing resources by SDN controller (for network) and Kubernetes (for server).
    \item We develop an online slice reconfiguration algorithm (OSRA) to re-configure and manage slices in resource-constrained scenarios.
    \item Compared to existing approaches, we observe that our mapping methodology meets the QoE requirements with high accuracy. It improves upon the existing approaches by approximately 98\% for bursty traffic variations. 
    \item Our algorithm has fast convergence and achieves low E2E delay violations for lower priority slices.
\end{itemize}

\section{Literature Review and Motivation}
\label{sec:litrev}
 
Reconfiguration of slices at core networks have been investigated in \cite{wei2020dynamic,wang2018fast}.  In \cite{wang2018fast}, the problem is solved using an L1- regularization of the reconfiguration cost. The authors in \cite{wei2020dynamic} take reconfiguration decisions such that over time, fewer resources are used. Since this is an optimization over time, they employ a reinforcement learning-based approach and learn allocation strategies that leave free resources as time progresses. Both of these works consider core networks where the traffic is much smoother compared to RAN counterpart. In addition, the authors assume that applications are aware of bandwidth and computing resources required to satisfy their QoE requirements and demand the same from the service provider. Thus, these works do not directly aim at guaranteeing E2E QoE which is the primary objective of creating slices. In contrast, in this paper we consider learning the relationship between E2E QoE and resources before configuring the slices. In addition, we provision reconfiguration of slices as application demands vary dynamically.

 As an example, consider a network, there might be existing slices which were configured based on service level agreements (SLAs) that meet the desired QoE requirements. Based on SLAs and type of applications, these slices can also be prioritized. For example, slices for best-effort and entertainment applications would have lower priority than mission critical slices serving, say health management. The following situations might arise in slice management: 
\begin{itemize}
    \item \textbf{SLA renegotiation}: There might be an increase in traffic flows in the existing slices, beyond agreed-upon SLAs. In this situation, users/businesses will have degraded experience and might want to renegotiate the SLAs. When free resources are not enough to satisfy the QoE, resources may be adjusted to support higher priority ones. In this scenario, we can use the learnt relationship between E2E QoE and resources for re-configuring the slices.
    \item \textbf{New slice creation}: New businesses might develop new applications that do not map to the existing slices which requires creating new slices. However, in this situation, there is no prior knowledge regarding the relationship between E2E QoE and resources for the new applications. The reconfiguration algorithm should be designed to adapt and learn the required relationship online.
\end{itemize} 

\begin{figure}[htp]
    \centering
    \includegraphics[width=0.79\linewidth]{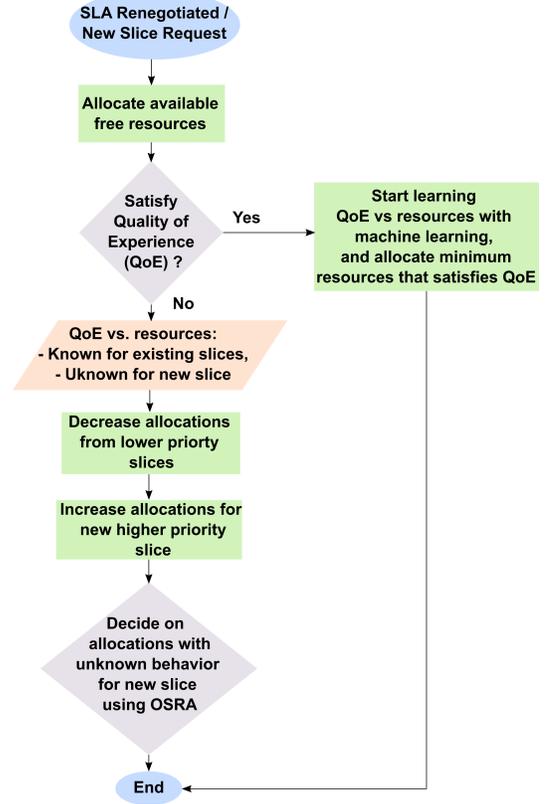}
    \caption{Conceptual Overview of our reconfiguration process with Online Slice Reconfiguration Algorithm (OSRA). We try to accommodate a new high-priority slice at the cost of degrading lower priority ones by adjusting resources. We need to learn the unknown behavior of the applications or estimate online during the reconfiguration process.}
    \label{fig:concept}
\end{figure}

Evidently, among the above two situations, new slice creation is the more generic case since it has more unknowns. Reconfiguration of existing slices might lead to violations of E2E QoE for existing applications which need to be minimized based on their priorities. In addition, reconfiguration time should be at minimum, so that new prioritized applications of the new slice can start their execution as early as possible. In this paper, we address the problem of E2E slice reconfiguration in RAN, such that E2E QoE of existing applications are minimally violated while new prioritised applications may be allocated with their slices in a timely manner. Owing to the monotonic relationship between resources and QoE metrics (allocating more resources cannot adversely affect QoE), we propose a reconfiguration algorithm based on gradient descent (greedy approach). The corresponding gradient for new slice or applications is estimated with help of E2E QoE metrics obtained at previous time steps of the descent algorithm. A conceptual overview of how our developed mechanism works is shown in Fig. \ref{fig:concept}.


\section{Model and Notation}
We consider a RAN where users are connected to a remote radio head (RRH) with a router and a Baseband Unit (BBU) (see Fig. \ref{fig:simsetup}). The BBU is equipped with Kubernetes or Docker system to manage containers (applications) that run on them. An SDN controller can configure the router, and users host different applications with QoE requirements served at the BBU. The SDN and Kubernetes create E2E slices. They can communicate with each other and share statistics related to QoE metrics required to decide the resources for a slice. 

In our model, we assume that applications in a slice send packets about application execution at an edge or cloud server. The packets carry the required data for executing the job. For example, an application might be an algorithm to sort an array where packets contain array data. In such applications, the type of application and the amount of data it sends determine the degree of processing required at the server. Thus, each application can have its traffic generation rates and processing demands. Further, depending on the type of protocol employed at MAC and transport layers, the delays experienced at the network sites may vary. We can characterize the E2E user experience by the delays experienced to obtain the result of an application request at the user sites. The SDN can collect such data as statistics for understanding E2E QoE and resource mappings. In addition, we may lose packets either at the network (considering a UDP based protocol) or at the server (due to overload or out of memory). In such cases, the user does not respond to the server leading to an unpleasant experience. To model this effect, we define E2E throughput as the percentage of successful requests. Thus, we use the two metrics, namely E2E delay and E2E throughput, for characterizing user experience.

Let $\mathbb{A} \in \{a_1,\dots,a_N\}$ denote the set of slices. Each slice $i$ demands a QoE depending on its requirements given by the tuple 
$q_i = (\tau_i,\rho_i)$, where $\tau_i$ denotes the desired E2E delay, $\rho_i$ denotes the desired E2E throughput. 
Let $j$ be a new slice which demands $q_j = (\tau_j,\rho_j)$ as E2E delay and throughput to be satisfied. The set $\mathbb{B} \subseteq \mathbb{A}$ denotes the set of slices or classes which have lower priority than $j$.

\begin{table}[!ht]
\caption{Notations used in this work.}
    \centering
    \begin{tabular}{|p{1.5cm}|p{6cm}|}
        \hline
        \multicolumn{1}{|c|}{\textbf{Notation}} & \multicolumn{1}{|c|}{\textbf{Description}}  \\
        \hline
        $\mathbb{A}$ & Set of Application Classes or Slices\\
        \hline
        $\mathbb{B}$ & Set of Slices which are of lower priority than Slice $j$\\
        \hline
        $G=(V,E)$ & Physical network graph. $V$ is set of vertices and $E$ is the set of edges/links. Routing decisions rely on link capacities.\\
        \hline
        \multicolumn{2}{c}{\textbf{QoE Parameters}}\\
        \hline
        ${\tau_i}$ & E2E delay requirement for slice $a_i\in \mathbb{A}$\\
        \hline
        ${\rho_i}$ & E2E throughput requirement for slice $a_i\in \mathbb{A}$\\
        \hline
        \multicolumn{2}{c}{\textbf{Slicing Decision Variables}}\\
        \hline
        $f_i^e$ & Flow for $a_i\in \mathbb{A}$ along edge $e\in E$ (implemented by SDN/Network Controller)\\
        \hline
        $\phi_i^c$ & Fraction of CPU core $c$ allocated at the server side  for $a_i\in \mathbb{A}$ (implemented by Kubernetes)\\
        \hline
        $\mathbf{s_i}$ & $\big[f_i^e,
        \phi_i^c\big]'$\\
        \hline
        \multicolumn{2}{c}{\textbf{Functions and Sets in Optimization}}\\
        \hline
         $\mathbb{C}$ & Constraint Set \\
        \hline
        $D_i()$ & Delay constraint function\\
        \hline
        $T_i()$ & Throughput constraint function \\
        \hline
        $\pi_i()$ & Penalty function for graceful degradation of constraint to be relaxed\\
        \hline
        $\Pi_\mathbb{C}(.)$ & Projection operator on $\mathbb{C}$.\\
        \hline
    \end{tabular}
    \label{notation_tab}
\end{table}

Typically, network slicing involves decisions regarding bandwidth slice, path/route and the  corresponding processor allocation at the edge/cloud server. Given a graph $G = (V,E)$, a slicing decision may be modeled as a tuple $s_i = (f_i^e,\phi_i^c)$, where $f_i^e \in [0,1]$ denotes the flow for $a_i$ along edge $e$. This decision on $f_i^e$ is taken by an SDN controller or network hypervisor, such as FlowVisor. In this work, routing decisions have not been considered due to the considered topology. However, the developed methodology can be extended to routing scenarios as well. $\phi_i^c \in [0,1]$ denotes the fraction of the processing for core $c$ at a server. This modeling allows us to capture parallalizable applications as well and its realistic implementations are feasible due to the advent of dockers and Kubernetes based server systems \cite{docker,kubernetes_2021}. 
In our experiments and in this paper, we assume that the server has sufficient RAM to at least execute an application. Thus, we have ignored the decisions for RAM in the paper. However, the developed model can be suitably extended to incorporate RAM allocations as well.
\subsection{QoE constraints}
Each of the two major E2E QoE parameters, namely delay and throughput can be captured via implicit functions of the form $D_i(f_i^e,\phi_i^c)$ and $T_i(f_i^e,\phi_i^c)$ for E2E delay and throughput respectively. Note that these functions are of stochastic nature and hence by $D_i()$, we mean a statistic of the delay like maximum delay or delays for 95\% of the requests. In this paper, we used maximum delay as the metric to capture worst case situations. Other QoE parameters can also be included to extend our model. The QoE functions discussed before are restricted by the following bounds:
\begin{align}
    D_i(f_i^e,\phi_i^c)&\leq \tau_i ~~~\forall ~i\\
    T_i(f_i^e,\phi_i^c) &\geq \rho_i ~~~\forall ~i
\end{align}
For predefined slices (network and computing resource allocation), the functions $D_i(.)$ and $T_i(.)$ are known and would be satisfying these constraints. For new applications of slice $j$, $D_j(.)$ and $T_j(.)$ are unknown.

\subsection{Link Capacity and Server Capacity}
Given a link $e\in E$ from the network graph $G$, the capacity of the link is bounded by $B(e)$. The bandwidth allocated to applications routed through this link should not exceed this capacity. In RAN, all applications are routed through the same link and the $f_i^e$ can be translated to normalized bandwidth allocated for each slice in the link. 
\begin{align}
    \sum_{i\in \mathbb{A}} f_i^e \leq 1 ~ ~~\forall e \in G
\end{align}
 The fraction of processing for a given core $c$ at a server should not exceed the maximum processing capability of the core. We consider a core as a unit processing element and all cores are identical. Thus, we must have
\begin{equation}
    \sum_{i \in \mathbb{A}} \phi_i^c \leq 1 ~~~\forall ~c \in C
\end{equation}
Here $C$ denotes the set of cores of a CPU in a server.
\begin{figure}
    \centering
    \includegraphics[width=0.7\linewidth]{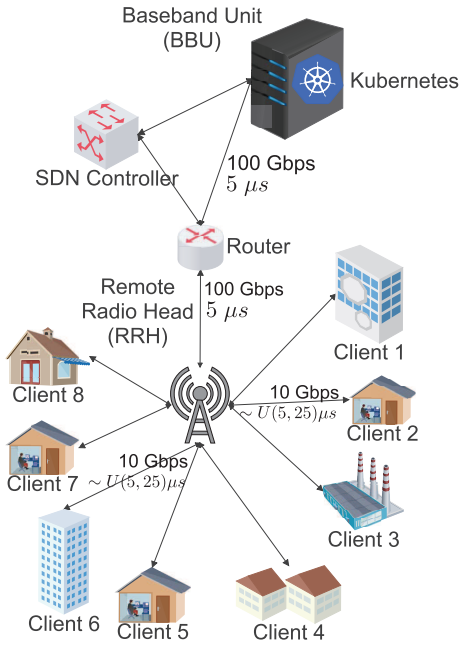}
    \caption{Simulation setting for E2E QoE constraints over RAN.}
    \label{fig:simsetup}
\end{figure}

\section{Problem Definition}
A new slice $j$ may have its delay and throughput constraints such that the available resources may not be able to satisfy the same. Then, we should reconfigure the existing slices to accommodate the new slice optimally. The notion of priority may be conceived as a mapping of the requested QoE parameters.  Based on the priority, the service provider may impose appropriate penalties, thereby allowing them to gracefully degrade (relax low priority constraints to make slicing feasible) the service. Since QoE parameters are to be relaxed, we define $\pi_i(D_i(),T_i())$ as the associated penalty function for slice $a_i$ that is applicable for violating the QoE constraints. We can design this function with some prioritization over the slices depending on payments received for executing them. An example penalty function (used in this paper) is:
\begin{equation}
\begin{split}
 \pi_i(D_i(.),T_i(.)) = \alpha_{i\tau}\max\{0,(D_i(.) - \tau_i)\}+\\
 \alpha_{i\rho}\max\{0,(\rho_i-T_i(.))\} ~ \forall ~ i\in \mathbb{B}   
\end{split}
\end{equation}
Here, $\alpha_i$ factors in the importance given to slice $i$ with respect to others. Quadratic penalties are applied, where the objective is to minimize the QoE violations. However, it is important to note that $D_j(.)$, $T_j(.)$ and thus $\pi_j(.)$ are unknown functions and have to be learnt on the fly while deciding on the allocation. Also, $\alpha_j>\alpha_i ~\forall ~i\in \mathbb{B}$.

The reconfiguration problem can be cast as: 
\begin{subequations}
\begin{gather}
    \min_{(f_i^e,\phi_i^c)}  \sum_{i\in \mathbb{B}}  \pi_i(D_i(.),T_i(.))+ \pi_j(D_j(.),T_j(.))     \label{grafit} \\
    \text{ s.t. }\sum_{i|a_i\in \mathbb{B}} f_i^e \leq 1 ~~\forall e \in G \label{gracon3} \\
    \sum_{i|a_i \in \mathbb{B}} \phi_i^c \leq 1 ~~\forall ~c \in C  \label{gracon4} \\
    f_i^e\geq0, ~ \phi_i^c\geq0 ~\forall ~ i,j
    \label{gracon5}
\end{gather}
\end{subequations}

The constraints \eqref{gracon3}, \eqref{gracon4}, \eqref{gracon5} can be combined to form a convex constraint set $\mathbb{C}$. We now devise an algorithm to learn and allocate resources to slice $j$.

\begin{figure*}[!ht]
    \centering
    \subfloat[]{\includegraphics[width=0.33\linewidth]{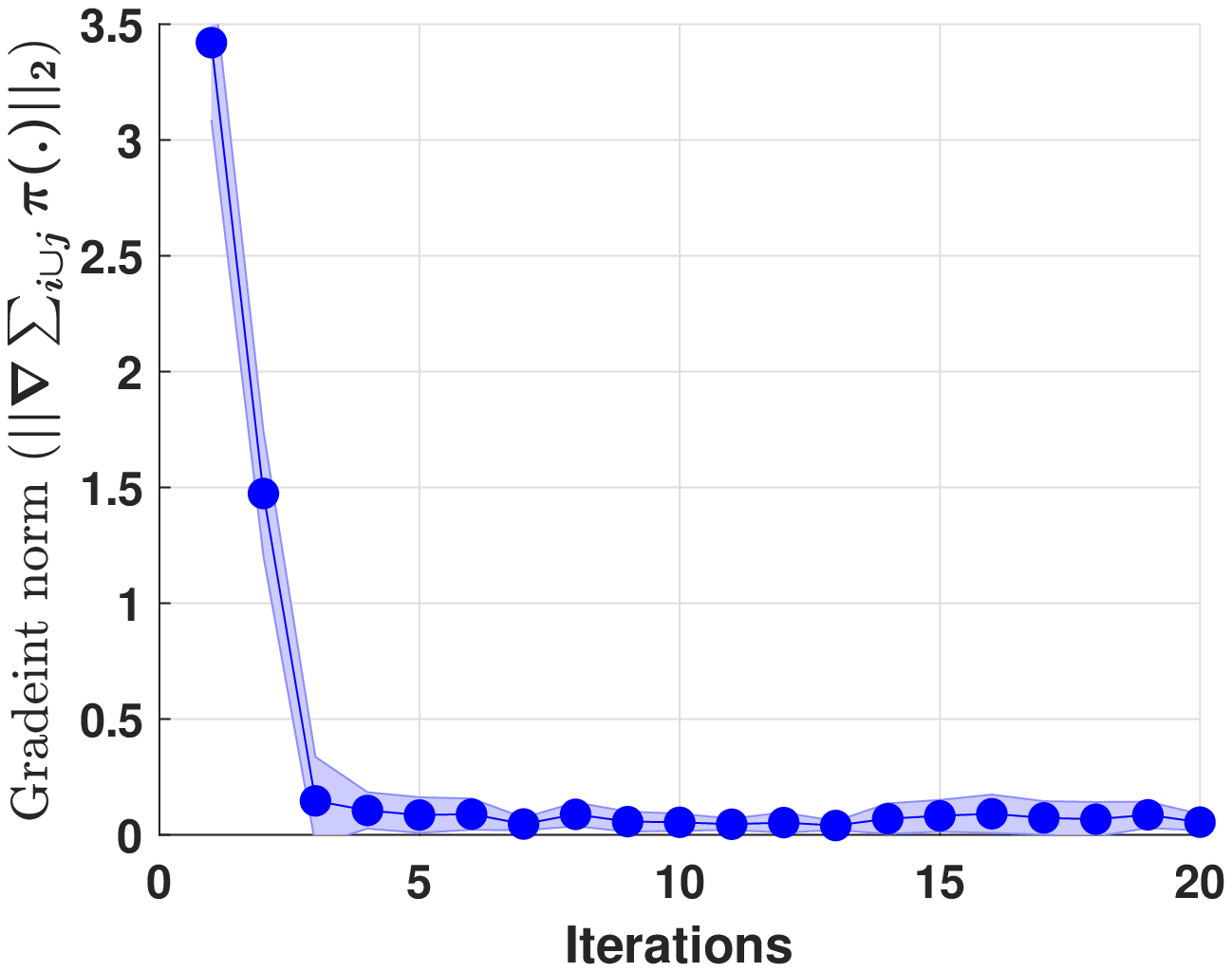}} \hfill
    \subfloat[]{\includegraphics[width=0.33\linewidth]{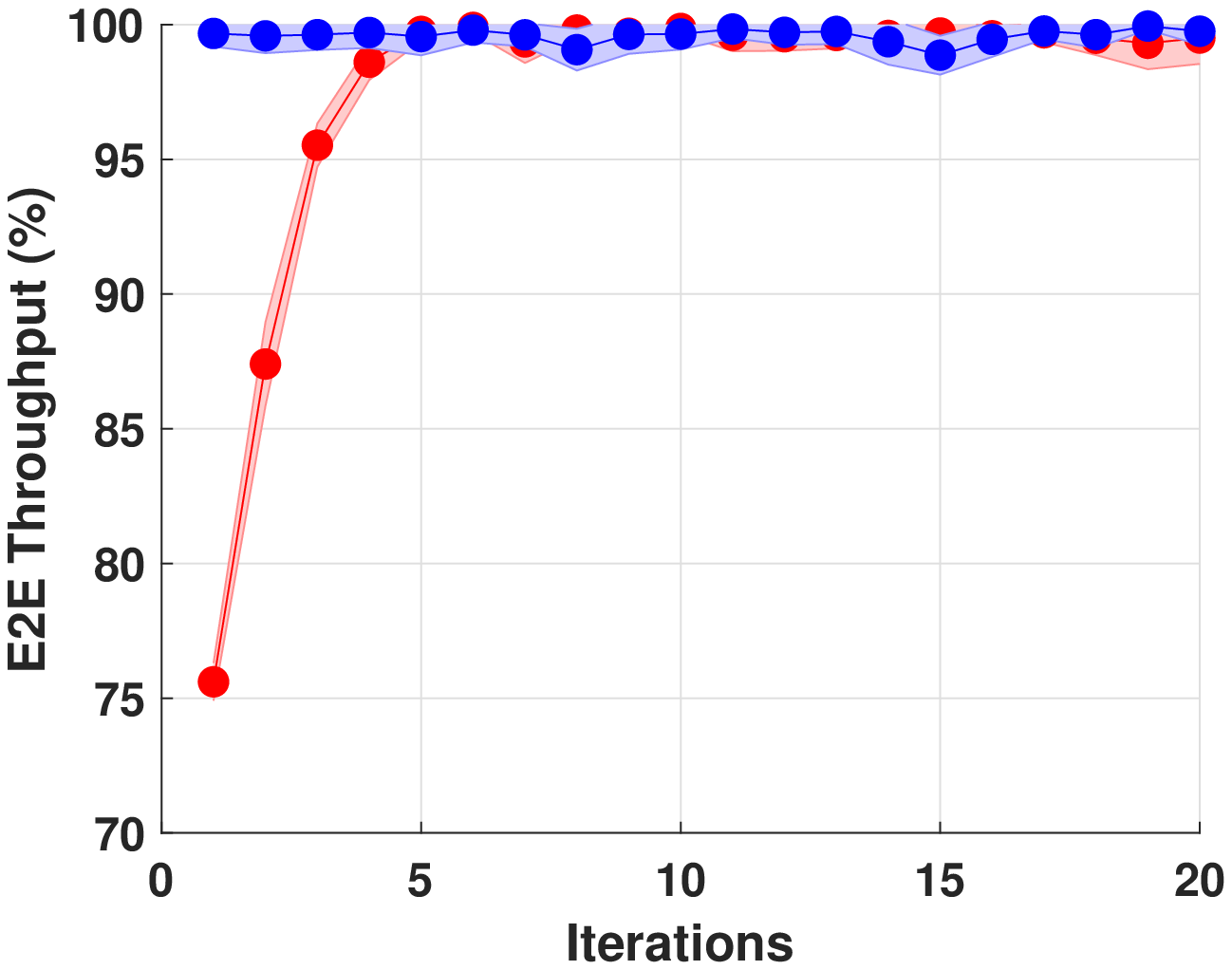}} \hfill
    \subfloat[]{\includegraphics[width=0.33\linewidth]{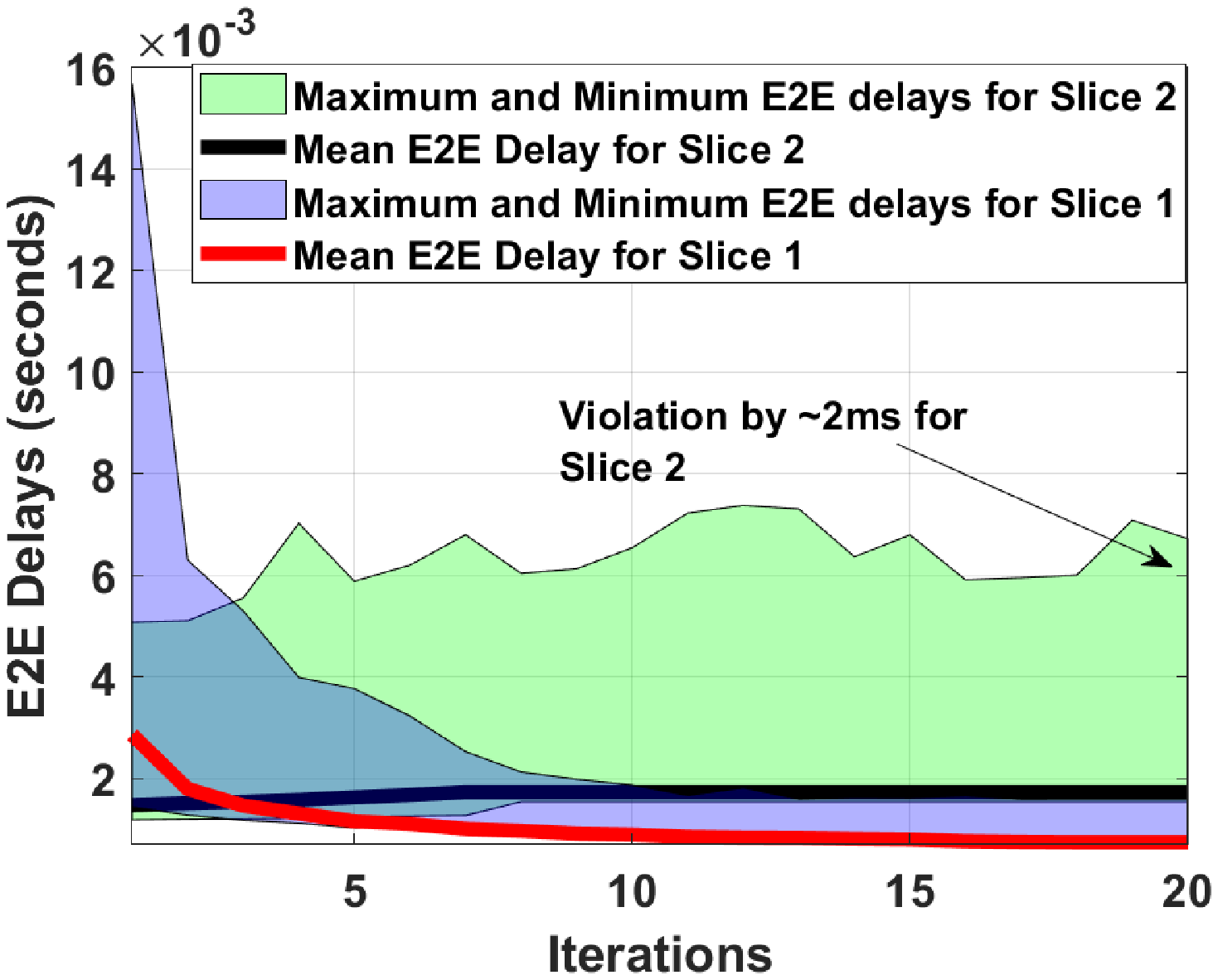}}
    \caption{Performance of OSRA with iterations and E2E QoE metrics: (a) convergence rate of the algorithm for 10 different seeds; (b) convergence of E2E throughput, and  (c) convergence of E2E delays of the two slices---the shaded region shows the variation of delays due to traffic variations with mean delays.}
    \label{fig:e2edelay}
\end{figure*}

\begin{figure*}[!ht]
    \centering
    \subfloat[]{\includegraphics[width=0.45\linewidth]{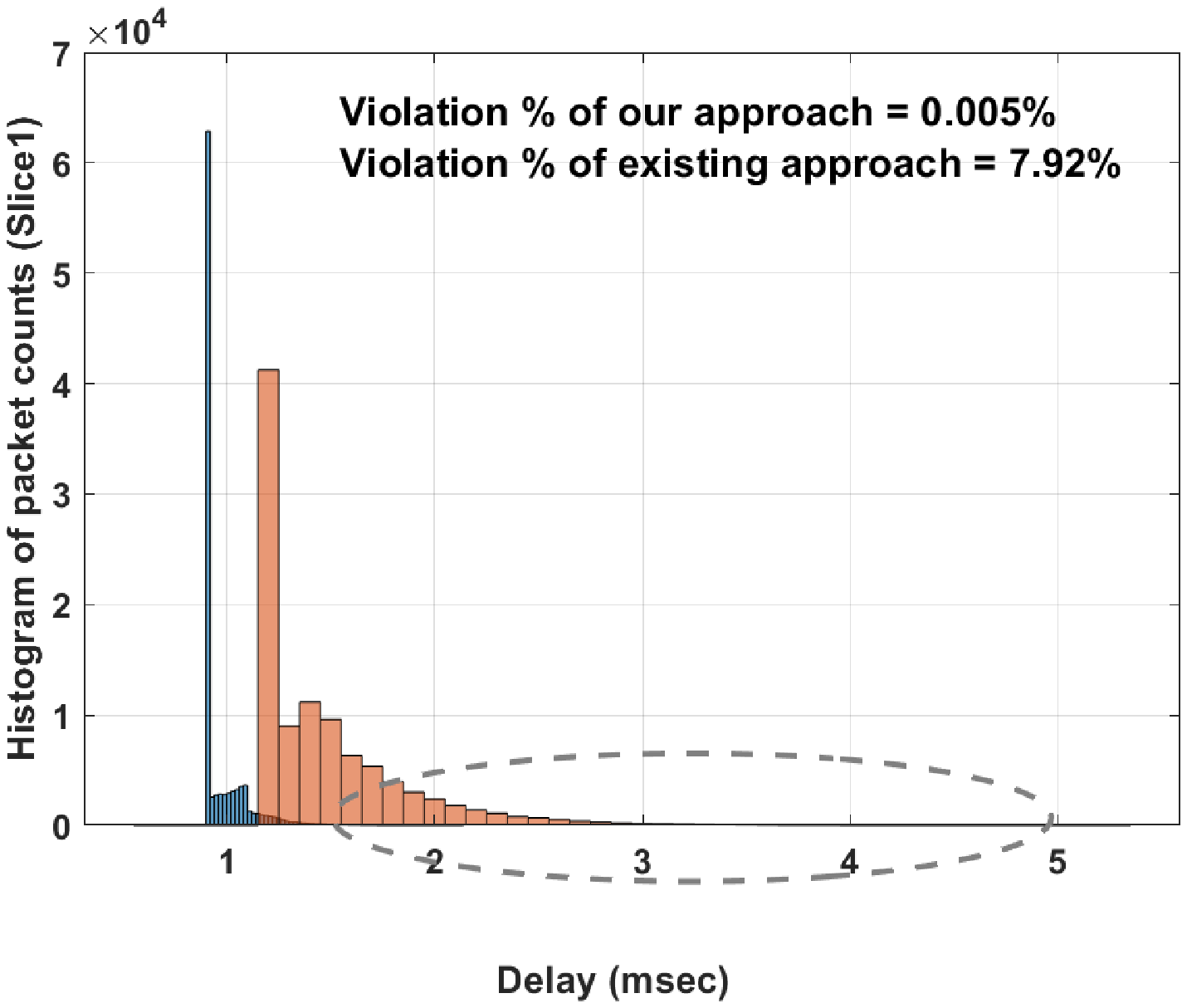}} \hfill
    \subfloat[]{\includegraphics[width=0.45\linewidth]{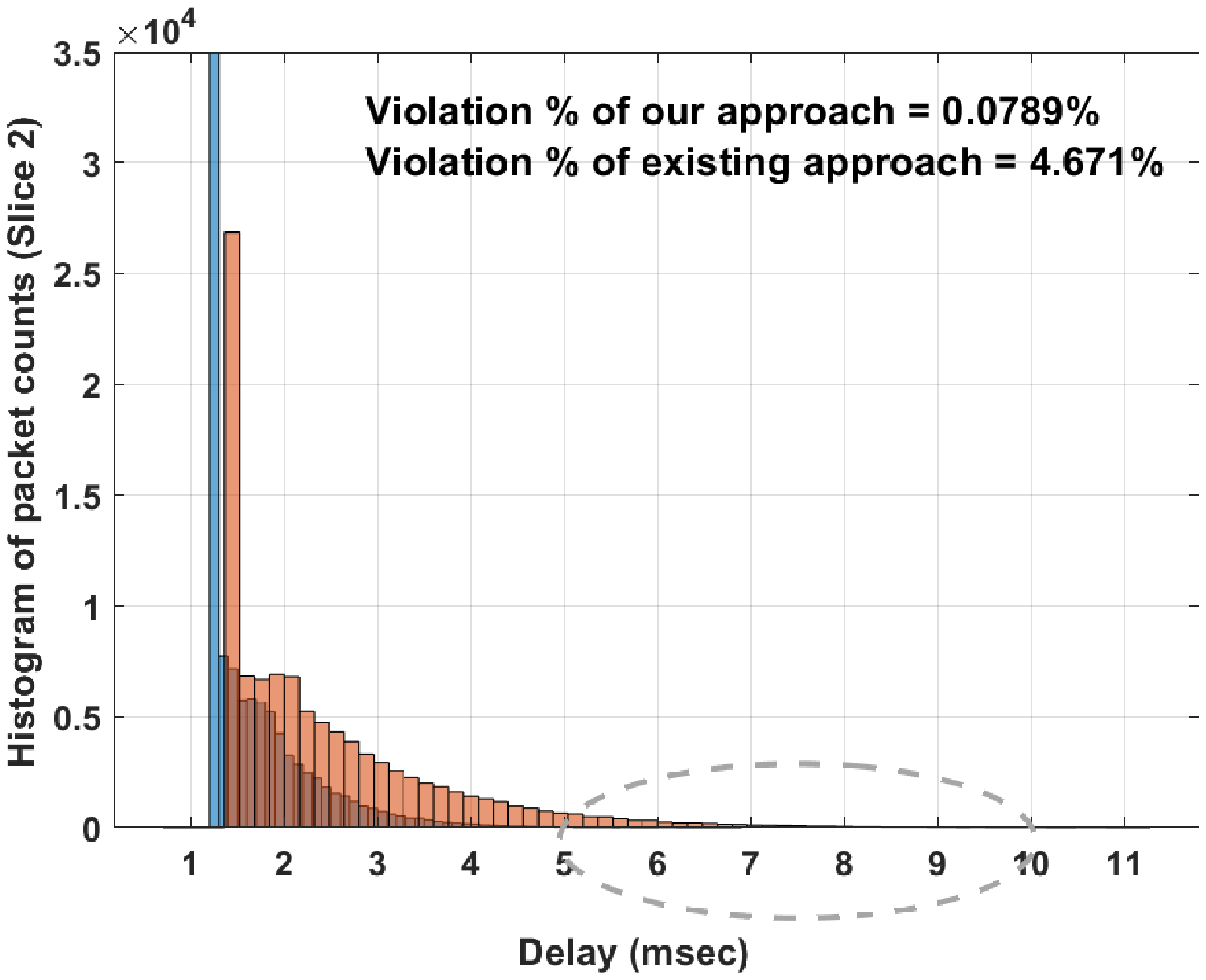}} 
    \caption{Comparison of OSRA with Slicing Approaches (known resource) \cite{wei2020dynamic,wang2018fast}(a) Delay Violations for Slice 1 with OSRA and Existing Slicing Approach; (b) Delay Violations  for Slice 2 with OSRA and Existing Slicing Approach}
    \label{fig:compare}
\end{figure*}

\section{Online Slice Reconfiguration Algorithm (OSRA)}
We use a gradient descent based approach, which is an iterative algorithm (\ref{alg:two}).  The idea is to decrease resources from lower priority slices and increase for the new one.

\RestyleAlgo{ruled}
\SetKwComment{Comment}{/* }{ */}
\SetKwInOut{KwIn}{Input}
\SetKwInOut{KwOut}{Output}
\begin{algorithm}[hbt!]
\caption{Reconfiguration Algorithm with Online Gradient Estimation}\label{alg:two}
\KwIn{Known Penalty Functions: $\pi_i(.) ~~\forall i$, Current Allocations:$\bm{s_i}(k)$}

\KwOut{Optimal Allocations: $ \bm{s_i^*} = (f_{i}^{*},\phi_{i}^{*}) ~~\forall i,j$}
\While{}{
Compute Gradients Existing Slices: $\nabla \pi_i(\bm{s_i}(k))$\;
Estimate Gradient for New/Unknown Slice:
$\nabla \pi_j(\bm{s_j}(k))$ using \eqref{eq:approx}\;
$\bm{s_i}(k+1) = \Pi_\mathbb{C}\Big(\bm{s_i}(k) - \eta_i(k) \nabla \pi_i(\bm{s_i}(k))\Big), \forall ~ i\in \mathbb{B}$\;
$\bm{s_j}(k+1) = \Pi_\mathbb{C}\Big(\bm{s_j}(k) + \frac{\sum_i\eta_i(k) \nabla \pi_i(\bm{s_i}(k))}{{\lVert \nabla \pi_j(\bm{s_j}(k) \rVert}}\Big)$\;
Stopping Criteria:
$\lVert \sum_{i}\eta_i(k) \nabla \pi_i(\bm{s_i}(k))\rVert \leq \epsilon$
}
\end{algorithm}

The nature of penalty functions is fundamental to decide the amount to be withdrawn or added to the new slice at each step. Let $k$ denote the iteration index. We may then denote the resources allocated for slice $i$ as $f_i^e(k)$ and $\phi_i^c(k)$ respectively. We drop the superscripts for ease of notation. At the next index, $k+1$, we move from the current point as:
\begin{subequations}
\begin{equation}
\bm{s_i}(k+1) = 
\begin{bmatrix}
f_i(k+1)\\
\phi_i(k+1)
\end{bmatrix}
=
\begin{bmatrix}
f_i(k) - \delta f_i(k)\\
\phi_i(k) - \delta \phi_i(k)
\end{bmatrix}
\end{equation}
\begin{equation}
\bm{s_j}(k+1) = 
\begin{bmatrix}
f_j(k+1)\\
\phi_j(k+1)
\end{bmatrix}
=
\begin{bmatrix}
f_j(k) +\sum_i\delta f_i(k)\\
\phi_j(k) + \sum_i \delta \phi_i(k)
\end{bmatrix}
\end{equation}
\end{subequations}
Then, we have:
\begin{subequations}
\begin{equation}
\pi_i(\bm{s_i}(k+1)) \approx \pi_i(\bm{s_i}(k)) - \delta \bm{s_i}(k)^T \nabla \pi_i(\bm{s_i}(k))
\end{equation}
\begin{equation}
\pi_j(\bm{s_j}(k+1)) \approx \pi_j(\bm{s_j}(k)) + \Big(\sum_i \delta \bm{s_i}(k)\Big)^T \nabla \pi_j(\bm{s_j}(k))
\end{equation}
\end{subequations}
Using the approach of gradient descent, if we choose
$
    \delta \bm{s_i}(k) = \eta_i(k)\nabla \pi_i(\bm{s_i(k)})
$,
we must have 
$
    \sum_i \delta \bm{s_i}(k) = \sum_i \eta_i(k)\nabla \pi_i(\bm{s_i}(k)) = -\eta_j(k) \nabla\pi_j(\bm{s_j}(k))  
$
or,
\begin{equation}
\label{eq:rate}
     \eta_j(k) =  - \frac{\sum_i \eta_i(k)\nabla \pi_i(\bm{s_i}(k))}{\|\nabla\pi_j(\bm{s_j}(k))\|}
\end{equation}
\eqref{eq:rate} gives us the step size for slice $j$ in terms of others. Its important to note that if $\nabla \pi_i(\bm{s_i}(k)), ~ \forall ~ i\in \mathbb{B}\cup\{j\}$ are linearly independent, the learning rates should be trivially zero. However, due to monotonic behavior of the delay and throughput functions with respect to resources, allocating more resources can never affect adversely. This, allows us to choose \eqref{eq:rate} as our step size update for slice $j$. Step sizes for other slices can be chosen using different techniques available for choosing step sizes in a descent algorithm.

As we have mentioned before, the function $\pi_j(.)$ is not known. However, for the reconfiguration algorithm to work, we need only the gradient of  $\pi_j(.)$ at a given point.  This may be approximated numerically by finite difference method which is a first order approximation of the gradient.
\begin{equation}
\label{eq:approx}
\nabla \pi_j(\bm{s_j}(k)) = \frac{\pi_j(\bm{s_j}(k)+\Delta) - \pi_j(\bm{s_j}(k)-\Delta)}{2\Delta} 
\end{equation}
With progress in time, more data points for $\pi_j(.)$ are obtained and $\pi_j(.)$ is gradually learnt (all points are kept in memory and one can use these points to learn the function using machine learning techniques). However, as evident from the approach, at each step we take an approximate (not accurate since the function is not known) greedy decision so that the QoE violations are appropriately penalized. In addition to aforementioned updates, it must be ensured that the solutions remain in the constraint set $\mathbb{C}$. Evidently $\mathbb{C}$ is a convex set and more specifically a simplex. We denote the projection operator on the set by $\Pi_\mathbb{C}()$ \cite{wang2013projection}. 
\begin{equation}
    \Pi_\mathbb{C}(\mathbf{x_0}) = \argmin_{\mathbf{x}\in \mathbb{C}} \frac{1}{2} \|\mathbf{x}-\mathbf{x_0}\|_2^2
\end{equation}
Algorithm \ref{alg:two} shows the reconfiguration algorithm resulting from the discussed steps. To implement Algorithm \ref{alg:two}, the SDN controller maintains learnt neural network models (as function approximators) for resource vs E2E QoE mappings for existing slices. The controller can duplicate applications and host the same at the server in order to probe and collect the E2E QoE metrics (similar to implementing a virtual network function). At each step of the algorithm, the new slice is allocated more resources. To calculate \eqref{eq:approx}, the SDN probes the network and server with $\Delta$ increments around the current allocation point $\bm{s_j}(k)$. Since the delays and throughput are stochastic, the probing is done multiple times and the relevant statistics (like average or max or any other percentile) depending on requirement is used to calculate \eqref{eq:approx}.


\section{Experimental Results}
We conduct multiple experiments using simulation environments. These experimental scenarios are motivated by a suggestion from an actual network operator (Telstra) in the deployment of 5G networks in Australia. We simulated the considered network in the OmNet++. Using this simulator, we present large-scale simulation studies.
 
Fig. \ref{fig:simsetup} shows the simulation scenario for our setup and.  In our setup, we had two existing slices and a new slice creation request was considered. The new slice is named Slice 1 with E2E delay and throughput requirements of 2 ms and 99.9\% (mimicing uRLLC requirements). Slice 2 (eMBB) and slice3 (best-effort) were the existing slices. Slice 2 was configured with requirements of  5 ms and 95\% while best-effort has no delay constraint but throughput of 100\%. Our results show the performance of OSRA on Slice 1 and Slice 2. The applications in these slices generate packets that carry requests for a server to perform a processing task. We generated the packets using a UDP burst generator (available in INET with OmNet++) with sizes $\sim U(20, 65535)$. The processing requirements of the tasks are in terms of million instructions (MI), and the server uses a MIPS instruction set. Slice 1 requested 5e4 MI while Slice 2 requested 8e4 MI. The server operates at 3e8 MIPS and has 2 cores. Fig. \ref{fig:simsetup} provides the network configuration parameters. We now evaluate the performance of our reconfiguration algorithm based on its convergence rate and compare it with the existing approaches.

\subsection{Performance on Reconfiguration Time}
Fig. \ref{fig:e2edelay} shows the performance of the developed reconfiguration algorithm with iterations. All figures show that the algorithm converges very within 5-6 steps due to the greedy approach. It is important to note that allocating more resources will always aid in satisfying QoE resources which indicates a monotonic behavior and is fundamental for the choice of our approach. We carried out each experiment over ten runs with different seeds for generating traffic, and the shaded areas show the variations. Fig. \ref{fig:e2edelay} shows the variations of E2E delays. The shaded areas show the deviation of maximum E2E delays with respect to the mean. Due to bursty nature of traffic, we observe that the maximum delay has large variations compared to the mean. The bold lines show the mean delays. 

Evidently with iterations, the mean delays for Slice 1 decrease at the cost of increase for Slice 2. Similar trend is seen in E2E throughput as well. While most packets were initially getting lost for Slice 1, with increase in iterations the E2E throughput increases. However, we observe that E2E throughput for Slice 2 remains relatively unaffected while there is a minor degradation in E2E delay performance. This is majorly due to the fact that number of dropped packets is more reliant on the bandwidth allocations and the limited buffer size of the routers than the limitations at the server side. From the server side, only failures such as out of memory/system crash can lead to loss of packets or requests. As seen from the obtained results, Slice 2 suffers a delay violation of approximately 2ms in order to accommodate Slice 1.



\subsection{Comparison with Existing Slicing Approaches}
As discussed in Section \ref{sec:introduction}, the existing slicing approaches assume that the resources required to satisfy the QoE requirements are known and hence are fed as demand inputs to their respective optimization formulations. One might assume this knowledge from known analytical relationships between delay and serving power. Hence to draw a comparison, we obtain the estimate of the required bandwidth and processing power required to obtain an E2E delay by considering separate M/M/1 queuing models at the server and the network sites. We use the arrival rates and serving rates from our simulation setup. As shown in Fig. \ref{fig:compare}, we plot the histogram of received packets and the E2E delays experienced by the packets returning from the edge server. As seen from the figures, the resources obtained from the analytical models violate delays by more than 98\%. This shows the efficacy of using the learning based approach towards creating and managing slices in the network.

\section{Conclusion}
In this paper, we addressed the problem of online reconfiguration of slices at the fronthaul segment of RAN. The bursty nature of traffic significantly affects the delays and throughput, which presents a challenge while re-configuring existing slices. A higher priority slice may be accommodated by gracefully degrading lower priority ones with the help of resource exchanges. We provide a gradient-based greedy algorithm to facilitate these exchanges. Further improvements with sophisticated variations of the algorithm may improve the presented results. We haven't considered jitter constraints in the paper, which might be of interest especially in RAN.

\bibliographystyle{IEEEtran}
\bibliography{ref}

\begin{thebibliography}{10}
\providecommand{\url}[1]{#1}
\csname url@samestyle\endcsname
\providecommand{\newblock}{\relax}
\providecommand{\bibinfo}[2]{#2}
\providecommand{\BIBentrySTDinterwordspacing}{\spaceskip=0pt\relax}
\providecommand{\BIBentryALTinterwordstretchfactor}{4}
\providecommand{\BIBentryALTinterwordspacing}{\spaceskip=\fontdimen2\font plus
\BIBentryALTinterwordstretchfactor\fontdimen3\font minus
  \fontdimen4\font\relax}
\providecommand{\BIBforeignlanguage}[2]{{%
\expandafter\ifx\csname l@#1\endcsname\relax
\typeout{** WARNING: IEEEtran.bst: No hyphenation pattern has been}%
\typeout{** loaded for the language `#1'. Using the pattern for}%
\typeout{** the default language instead.}%
\else
\language=\csname l@#1\endcsname
\fi
#2}}
\providecommand{\BIBdecl}{\relax}
\BIBdecl

\bibitem{sachs20185g}
J.~Sachs, G.~Wikstrom, T.~Dudda, R.~Baldemair, and K.~Kittichokechai, ``5g
  radio network design for ultra-reliable low-latency communication,''
  \emph{IEEE network}, vol.~32, no.~2, pp. 24--31, 2018.

\bibitem{letaief2019roadmap}
K.~B. Letaief, W.~Chen, Y.~Shi, J.~Zhang, and Y.-J.~A. Zhang, ``The roadmap to
  6g: Ai empowered wireless networks,'' \emph{IEEE Communications Magazine},
  vol.~57, no.~8, pp. 84--90, 2019.

\bibitem{Yourguid58:online}
``Your guide to 5g network automation and zero-touch,''
  \url{https://www.ericsson.com/en/blog/2020/11/your-guide-to-5g-network-automation-and-zero-touch},
  (Accessed on 10/06/2021).

\bibitem{chartsias2017sdn}
P.-K. Chartsias, A.~Amiras, I.~Plevrakis, I.~Samaras, K.~Katsaros,
  D.~Kritharidis, E.~Trouva, I.~Angelopoulos, A.~Kourtis, M.~S. Siddiqui
  \emph{et~al.}, ``Sdn/nfv-based end to end network slicing for 5g multi-tenant
  networks,'' in \emph{2017 European Conference on Networks and Communications
  (EuCNC)}.\hskip 1em plus 0.5em minus 0.4em\relax IEEE, 2017, pp. 1--5.

\bibitem{kubernetes_2021}
\BIBentryALTinterwordspacing
``Pod priority and preemption,'' Jul 2021. [Online]. Available:
  \url{https://kubernetes.io/docs/concepts/scheduling-eviction/pod-priority-preemption/}
\BIBentrySTDinterwordspacing

\bibitem{docker}
\BIBentryALTinterwordspacing
``Build docker kubernetes-ready applications on your desktop.'' [Online].
  Available: \url{https://www.docker.com/products/kubernetes}
\BIBentrySTDinterwordspacing

\bibitem{she2019cross}
C.~She, Y.~Duan, G.~Zhao, T.~Q. Quek, Y.~Li, and B.~Vucetic, ``Cross-layer
  design for mission-critical iot in mobile edge computing systems,''
  \emph{IEEE Internet of Things Journal}, vol.~6, no.~6, pp. 9360--9374, 2019.

\bibitem{wang2018fast}
G.~Wang, G.~Feng, T.~Q. Quek, and S.~Qin, ``On fast slice reconfiguration,'' in
  \emph{2018 IEEE Global Communications Conference (GLOBECOM)}.\hskip 1em plus
  0.5em minus 0.4em\relax IEEE, 2018, pp. 1--7.

\bibitem{wei2020dynamic}
F.~Wei, G.~Feng, Y.~Sun, Y.~Wang, and Y.-C. Liang, ``Dynamic network slice
  reconfiguration by exploiting deep reinforcement learning,'' in \emph{ICC
  2020-2020 IEEE International Conference on Communications (ICC)}.\hskip 1em
  plus 0.5em minus 0.4em\relax IEEE, 2020, pp. 1--6.

\bibitem{rossi2020hierarchical}
F.~Rossi, V.~Cardellini, and F.~L. Presti, ``Hierarchical scaling of
  microservices in kubernetes,'' in \emph{2020 IEEE International Conference on
  Autonomic Computing and Self-Organizing Systems (ACSOS)}.\hskip 1em plus
  0.5em minus 0.4em\relax IEEE, 2020, pp. 28--37.

\bibitem{hirai2020automated}
S.~Hirai, T.~Tojo, S.~Seto, and S.~Yasukawa, ``Automated provisioning of
  cloud-native network functions in multi-cloud environments,'' in \emph{2020
  6th IEEE Conference on Network Softwarization (NetSoft)}.\hskip 1em plus
  0.5em minus 0.4em\relax IEEE, 2020, pp. 1--3.

\bibitem{Microsof16:online}
``Microsoft powerpoint - tf1\_1704\_tam\_transport network slicing\_2,''
  \url{https://sagroups.ieee.org/1914/wp-content/uploads/sites/92/2017/04/tf1_1704_Tam_Transport-Network-Slicing_2.pdf},
  (Accessed on 10/05/2021).

\bibitem{wang2013projection}
W.~Wang and M.~A. Carreira-Perpin{\'a}n, ``Projection onto the probability
  simplex: An efficient algorithm with a simple proof, and an application,''
  \emph{arXiv preprint arXiv:1309.1541}, 2013.

\end{thebibliography}

\end{document}